\documentclass[12pt]{article}
\begin{document}
\begin{flushright}
OHSTPY-HEP-T-98-028 \\
hep-th/9811254
\end{flushright}
\vspace{20mm}
\begin{center}
{\LARGE The Light-Cone Vacuum in
1+1 Dimensional Super-Yang-Mills Theory}
\\
\vspace{20mm}
{\bf F.Antonuccio${}^{(1)}$, O.Lunin${}^{(1)}$,
  S.Pinsky${}^{(1)}$,
and
S.Tsujimaru${}^{(2)}$}\\
\vspace{4mm}
{\em ${}^{(1)}$Department of Physics,\\ The Ohio State University,\\
Columbus,
OH 43210, USA\\
\vspace{4mm} and \\
\vspace{4mm}
${}^{(2)}$Max-Planck-Institut f\"{u}r Kernphysik, \\ 69029 Heidelberg,
Germany}
\end{center}
\vspace{10mm}

\begin{abstract}
The Discrete Light-Cone Quantization (DLCQ) of a supersymmetric $SU(N)$
gauge theory
in 1+1 dimensions is discussed, with particular emphasis given to 
the inclusion of all dynamical zero modes. 
Interestingly, the notorious `zero-mode
problem' is now
tractable because of special supersymmetric cancellations. In particular,
we show that
anomalous zero-mode contributions to the currents are absent, in contrast
to what is
observed in the non-supersymmetric case. We find that the 
supersymmetric partner of
the gauge
zero mode is the diagonal component of the fermion zero mode. 
An analysis of the vacuum
structure is
provided and it is shown that the 
inclusion of zero modes is crucial
for probing
the phase properties of the vacua. In particular,
we find that the ground state energy is
zero and
$N$-fold degenerate, and thus consistent with unbroken supersymmetry. We
also show that the inclusion of zero modes for the light-cone supercharges
leaves the supersymmetry algebra unchanged.
Finally, we remark that the dependence of the light-cone Fock vacuum in
terms of the gauge zero  
is unchanged in the presence of matter fields.

\end{abstract}
\newpage
\section{Introduction}
A possibly surprising outcome of recent developments in string/M theory
are the proposed connections between non-perturbative
objects in string theory, and supersymmetric
gauge theories in low dimensions \cite{bfss97, dvv97}.
It is therefore of interest to study directly the non-perturbative
properties of super-Yang-Mills theories in various dimensions.

Recently, a class of 1+1 \cite{sakai95, alp98, alpII98, alp99}
and 2+1 \cite{alp99b} dimensional super Yang-Mills theories has been
studied using a supersymmetric form of Discrete Light-Cone Quantization
(SDLCQ) .
This formulation has the  advantage of preserving
supersymmetry after discretizing momenta, and admits
a very natural and straightforward algorithm for
extracting numerical bound state masses and wave functions
\cite{pb85, bpp98}.
Although a technical necessity, the omission
of zero-momentum modes in these numerical computations
raises many doubts about the consistency of such a
quantization scheme.
Little is in fact known about the precise effects of dropping the
zero-momentum mode at finite
compactification radius,
but it is generally believed that such effects disappear in the
decompactified limit \cite{alp99}.

There are instances, however, when we would like
to know the measurable effects of
a finite spatial compactification \cite{dvv97}.
In this work,
we will deal with measurable effects
that reflect the spatial compactification induced
by DLCQ. This is accomplished by explicitly
incorporating all the gauge zero mode degrees of freedom
in the DLCQ formulation of a supersymmetric gauge theory.
It turns out that this is tantamount to including a quantum mechanical
degree of freedom corresponding to
`quantized electric flux' around the compact direction\footnote{
There is also a fermion degree of freedom by virtue of supersymmetry.}.
 The supersymmetric partner of
this gauge degree of freedom is the diagonal component 
of the fermion zero mode.
The implications of this on the vacuum structure of the theory is
discussed.

The supersymmetric gauge theory we consider may be
obtained by dimensionally reducing ${\cal N} =1$ super-Yang-Mills
from 2+1 to 1+1 dimensions \cite{sakai95}.
The DLCQ formulation of this theory
consists of an adjoint scalar field (represented as a
$N \times N$ Hermitian matrix field), a corresponding
adjoint fermion field, and
several zero-mode (or quantum mechanical) degrees of freedom to be discussed
later.  To maintain supersymmetry one must impose
periodic
boundary conditions, so all the color degrees of freedom\footnote{
i.e. arising from the the generators of SU(N).} of the
fermion and boson fields will have zero modes.
In addition, periodic boundary conditions prevent us from
adopting the light-cone gauge, $A^+=0$, so we choose
instead the gauge $\partial_-A^+=0$, which allows $A^+$ to have a zero mode.
In addition, there are large gauge transformations, a Weyl
transformation, and a color permutation symmetry which
must be taken into account when constructing physical states of
the theory.
We briefly discuss this procedure.

In this work, we concentrate on the effect of including
the quantum mechanical degree of freedom represented
by the gauge zero mode. This zero mode corresponds to
a quantized color electric flux that circulates around the compact direction
$x^-$.  The problems associated with this zero mode have already been studied
in two dimensional gauge theories involving
adjoint scalars, and theories with adjoint fermions
\cite{kpp94,pk96,mrp97,kall,pin97a}. The consequences
of including these modes are quite drastic. These theories are known to
possess anomalies in the diagonal components of the current,
and therefore the charges in these theories
are time dependent. This makes it difficult -- if not impossible -- to
define a consistent theory. In contrast, owing to special
supersymmetric cancellations between boson and fermion currents,
no such anomalies arise in the supersymmetric theory studied here,
and so a DLCQ formulation becomes sensible and tractable.

In general, one finds a contribution after normal ordering the
Hamiltonian that is a function only of
the gauge zero-mode. This term acts as a vacuum potential and
leads to a non-zero vacuum energy.  When the gauge theory
without matter fields is solved, however, the only degree of freedom is the
quantum mechanical gauge zero mode, in which the vacuum potential plays no
role. The ground state energy is thus zero.
However, this simple picture of the vacuum may be drastically
altered if we consider the addition of matter.
For the supersymmetric case studied here, we show that
there is no vacuum potential, and that the ground state has zero energy
{\em even in the presence of matter fields.}

We find that the supersymmetric partner of the gauge zero mode is given
in terms of a diagonal component of the fermion zero mode, and therefore
the inclusion of fermion zero modes are necessary for a consistent 
treatment of the theory. Since the fermion zero modes commute
with the Hamiltonian they may be used to generate additional vacua. We
find that there are in fact
$N$ vacua consistent with the $SU(N)/Z_N$ reduced gauge invariance of the
theory.

This paper is organized as follows. In Section 2, we briefly
describe the DLCQ procedure of the 1+1 dimensional supersymmetric
Yang-Mills theory in the modified light-cone gauge.
In Section 3, the point splitting regularization designed to
preserve symmetry under large gauge transformations
is applied to the current operator. In Section 4,
we discuss the vacuum structure of the theory by deriving
the quantum mechanics of the zero modes.
We conclude in Section 5 with a brief discussion.

\section{Gauge Fixing in DLCQ}
We consider the supersymmetric Yang-Mills theory in
1+1 dimensions \cite{fer65} which is described by
the action
\begin{equation}
S  =  \int d^2 x \hspace{1mm}
 \mbox{tr} \left(-\frac{1}{4} F_{\mu \nu}F^{\mu \nu}
+\frac{1}{2}D_\mu \phi D^\mu \phi +
i \bar{\Psi}\gamma^\mu D_{\mu}\Psi -2ig\phi
\bar{\Psi}\gamma_5\Psi \right),
\end{equation}
where $D_\mu=\partial_\mu+ig[A_\mu, \cdot \hspace{1mm} ]$ and
$F_{\mu\nu}=\partial_\mu A_\nu -\partial_\nu A_\mu
+ig [A_\mu, A_\nu]$. All fields are in the adjoint
representation of the gauge group SU(N). A
convenient  representation of the gamma matrices is $\gamma^0=\sigma^2$,
$\gamma^1=i\sigma^1$ and $\gamma^5=\sigma_3$ where $\sigma^a$ are
the Pauli matrices. In this representation the Majorana spinor
is real. We use the matrix notation for $SU(N)$ so that $A^\mu_{ij}$ and
$\Psi_{ij}$ are $N\times N$ traceless matrices.

We now introduce the light-cone coordinates
$x^{\pm}=\frac{1}{\sqrt 2}(x^0 \pm x^1)$.
The longitudinal coordinate $x^-$ is compactified
on a finite interval $x^-\in [-L, L]$ \cite{my76, pb85}
and we impose periodic boundary conditions on all fields
to ensure unbroken supersymmetry.

The light-cone gauge $A^+=0$ can not be used in a finite
compactification radius,
but the modified condition $\partial_-A^+=0$ \cite{kpp94}
is consistent with the light-like  compactification.
We can make a
global rotation in color space so that
the zero mode is diagonalized $A^+_{ij}(x^+)=v_i(x^+)\delta_{ij}$ with
$\sum_i v_i =0$ \cite{kpp94}. The gauge zero modes correspond to a
(quantized) color
electric flux loops around the compactified space.
The modified light-cone gauge is not a complete gauge fixing. We still have
large gauge transformations preserving the gauge condition $\partial_-A^+=0$.
There are two types of such transformations \cite{lnt94,lst95}: displacements
$T_D$ and central conjugations $T_C$. Their actions on the physical fields of
the theory and complete gauge fixing will be discussed in the end of this
section. Now we just mention that being discrete transformations, $T_D$ and
$T_C$ don't affect quantization procedure.

The quantization in the light--cone gauge with or without dynamical $A^+$ is
widely explored in the literature \cite{pk96, mrp97, sakai95, alpII98, bpp98},
here we provide  only the results which are useful for later purposes.
The quantization proceeds in two steps. First, we must resolve the
constraints to eliminate the redundant degrees of freedom.
There are two constraints in the theory,
\begin{eqnarray}\label{constr1}
&&-D_-^2 A^- =gJ^+, \\
\label{constr2}
&&\sqrt{2} i D_- \chi=g[\phi, \psi],
\end{eqnarray}
where $\Psi\equiv (\psi, \chi)^{{\rm T}}$ and
the current operator is
\begin{equation}\label{current}
J^+(x)=\frac{1}{i}[\phi(x), D_-\phi(x)]-
\frac{1}{\sqrt 2}\{\psi(x), \psi(x)\}.
\end{equation}

Different components of (\ref{constr1}), (\ref{constr2}) play different roles
in the theory. First we look at diagonal zero modes of these equations.
The diagonal zero
mode of (\ref{constr2}) gives us constraints on the physical fields:
\begin{equation}\label{fermConstr}
[\phi,\psi]^0_{ii}=0.
\end{equation}
There is no sum over $i$ in above expression. As one can see this constraint
leads to decoupling of $\stackrel{0}{\chi}_{ii}$, this field plays the role
of Lagrange multiplier for above condition. The same is true for
${\stackrel{0}{A}}^-_{ii}$, the corresponding constraint is
$\stackrel{0}{J}_{ii}=0$.
The reason we treated the diagonal zero modes of (\ref{constr1}) and
(\ref{constr2}) separately is that for all other modes the $D_-$ operator is
invertible and instead of constraints on physical fields $\psi$ and $\phi$
one gets expressions for non-dynamical ones:
\begin{equation}
A^-=-\frac{g}{D_-^2}J^+,\qquad
\chi=\frac{g}{i\sqrt{2}}\frac{1}{D_-}[\phi, \psi].
\end{equation}

The next step is to derive the commutation relations
for the physical degrees of freedom.
As in the ordinary quantum mechanics, the zero mode
$v_i$ has a  conjugate momentum  $p=2L\partial_+v_i$ and
the commutation relation is $[v_i, p_j]=i\delta_{ij}$.
The off--diagonal components of the scalar field
are complex valued operators with
$\phi_{ij}=(\phi_{ji})^{\dagger}$. The canonical momentum
conjugate to $\phi_{ij}$ is $\pi_{ij}=(D_-\phi)_{ji}$.
They satisfy the canonical commutation relations \cite{pk96,sakai95}
\begin{equation}
[\phi_{ij}(x), \pi_{kl}(y)]_{x^+=y^+}=
[\phi_{ij}(x), D_-\phi_{lk}(y)]_{x^+=y^+}=\frac{i}{2}
(\delta_{ik}\delta_{jl}-\frac{1}{N}\delta_{ij}\delta_{kl})
\delta(x^--y^-).
\end{equation}
On the other hand, the quantization of the
diagonal component $\phi_{ii}$ needs care. As mentioned in
\cite{pk96}, the zero mode of $\phi_{ii}$, the mode independent
of $x^-$, is not an independent degree of freedom but
obeys a certain constrained equation \cite{my76, pk96, kall}.
Except the zero
mode, the commutation relation is canonical
\begin{equation}
[\phi_{ii}(x), \partial_-\phi_{jj}(y)]_{x^+=y^+}=
\frac{i}{2}(1-\frac{1}{N})\delta_{ij}
\left[\delta(x^--y^-)-\frac{1}{2L}\right].
\end{equation}
The commutator of diagonal and non-diagonal elements of $\phi$ vanishes.
The canonical anti-commutation relations for fermion fields are \cite{sakai95}
\begin{equation}
\{\psi_{ij}(x), \psi_{kl}(y)\}_{x^+=y^+}=
\frac{1}{\sqrt 2}
\delta(x^--y^-)(\delta_{il}\delta_{jk}-\frac{1}{N}\delta_{ij}\delta_{kl}).
\end{equation}
There are two differences between this expression and one from \cite{sakai95}.
First one is technical: we consider commutators for $SU(N)$ group, this gives
$1/N$ term. Second difference is that unlike \cite{sakai95} we include zero
modes in the expansion of $\psi$, we also include such modes in non-diagonal
elements of $\phi$.

Finally we return to the problem of complete gauge fixing. The actions of
$T_D$ and $T_C$ on physical fields are given by \cite{lnt94,lst95}:
\begin{eqnarray}
T_D:&& v_i(x^+)\rightarrow v_i(x^+)+\frac{n_i\pi}{gL},\qquad
       n_i\in {\bf {\rm Z}}, \qquad \sum n_i=0,\nonumber\\
 && \psi_{ij}\rightarrow \exp(\frac{\pi i(n_i-n_j)x^-}{L})\psi_{ij},\qquad
       \phi_{ij}\rightarrow \exp(\frac{\pi i(n_i-n_j)x^-}{L})\phi_{ij};\\
T_C:&& v_i(x^+)\rightarrow v_i(x^+)+\frac{\nu_i\pi}{gL},\qquad
       \nu_i=n(1/N-\delta_{iN}), \nonumber\\
 && \psi_{ij}\rightarrow \exp(\frac{\pi i(\nu_i-\nu_j)x^-}{L})\psi_{ij},\qquad
    \phi_{ij}\rightarrow \exp(\frac{\pi i(\nu_i-\nu_j)x^-}{L})\phi_{ij}.
\end{eqnarray}
There are also permutations of the color basis $i\rightarrow P(i)$ which
leave the theory invariant.
These symmetries preserve the gauge condition $\partial_-A^+=0$, but two
configurations related by $T_D$, $T_C$ or $P$ are equivalent. To fix the gauge
completely one therefore considers $v_i$ only in the fundamental domain,
other regions related with this domain by $T_D$, $T_C$ or $P$ give gauge
``copies" of it \cite{gri78}. The easiest thing to do is to describe the
boundaries of fundamental domain imposed by displacements $T_D$:
$-\frac{\pi}{2gL}<v_i<\frac{\pi}{2gL}$. The invariance under $T_C$
limits this region even more, but since we will not need the explicit form of
fundamental domain, we do not discuss such limits for $SU(N)$. For the
simplest case of $SU(2)$ the fundamental domain is given by
$0<v_1=-v_2<\frac{\pi}{2gL}$, the result for $SU(3)$ can be found in
\cite{lst95}. The $P$ symmetries do not respect the fundamental domain,
so they are not symmetries of gauge fixed theory. However there is one
special transformation among $P$ which being accompanied with combination of
$T_D$ and $T_C$ leaves fundamental domain invariant. Namely if $R$ is cyclic
permutation of color indexes then there exists a combination $T$ of $T_D$ and
$T_C$ such that $S=TR$ is the symmetry of gauge fixed theory. The explicit
form of $T$ depends on the rank of the group, for $SU(2)$ and $SU(3)$ it may
be found in \cite{lst95}. The operator $S$ satisfies the condition $S^N=1$
and it was used in classifying the vacua \cite{lst95,pin97a}.

\section{Current Operators}
The resolution of the Gauss-law constraint  (\ref{constr1})
is a necessary step for obtaining
the light-cone Hamiltonian.
The expression for the current operator is,
however, ill--defined unless an appropriate definition
is specified,  since
the operator products are defined at the same point.
We shall use the point--splitting regularization
which respects the symmetry of the theory under the
large gauge transformation.

To simplify notation it is convenient to introduce the dimensionless
variables $z_i=Lgv_i/\pi$ instead of quantum mechanical coordinates $v_i$
describing $A^+$.
The mode--expanded fields at the light-cone time $x^+=0$ are
\begin{eqnarray}
\phi_{ij}(x)&=&\frac{1}{\sqrt{4\pi}}
\left( \sum_{n=0}^{\infty} a_{ij}(n)u_{ij}(n) {\rm e}^{-ik_n x^-}
+\sum_{n=1}^{\infty} a^{\dagger}_{ji}(n) u_{ij}(-n)
{\rm e}^{ik_n x^-}\right), \qquad i\ne j,\nonumber \\
\phi_{ii}(x)&=&\frac{1}{\sqrt{4\pi}}
\sum_{n=1}^{\infty}\frac{1}{\sqrt n}\left(a_{ii}(n) {\rm e}^{-ik_n x^-}
+ a^{\dagger}_{ii}(n) {\rm e}^{ik_n x^-}\right), \nonumber \\
\psi_{ij}(x)&=&
\frac{1}{2^{\frac{1}{4}}\sqrt{2L}}
\left( \sum_{n=0}^{\infty}b_{ij}(n) {\rm e}^{-ik_n x^-}
+\sum_{n=1}^{\infty} b^{\dagger}_{ji}(n) {\rm e}^{ik_n x^-}\right),
\end{eqnarray}
where  $k_n=n\pi/L$, $u_{ij}(n)=1/\sqrt{\vert n-z_i+z_j \vert}$
\footnote{$u_{ij}(n)$ is
well-defined in the
fundamental domain. Similarly,
$(D_-)^2$ in the Gauss-law constraint
have no zero modes in this domain.}.
The (anti)commutation
relations for Fourier modes are found in \cite{pk96, mrp97}
and in our notation they take the form
\begin{eqnarray}\label{ccr}
&&[a_{ij}(n), a^{\dagger}_{kl}(m)]={\rm sgn}(n+z_j-z_i)\delta_{n, m}
(\delta_{ik}\delta_{jl}-\frac{1}{N}\delta_{ij}\delta_{kl}),\nonumber \\
&&\{b_{ij}(n), b^{\dagger}_{kl}(m)\}=\delta_{n, m}
(\delta_{ik}\delta_{jl}-\frac{1}{N}\delta_{ij}\delta_{kl})
\end{eqnarray}
The zero modes in above relations deserve special consideration. Although we
formally wrote them as $a_{ij}(0)$ and $b_{ij}(0)$, these modes also
act as creation operators because the conjugation of zero mode gives another
zero mode:
\begin{equation}
a_{ij}^\dagger(0)=a_{ji}(0), \qquad b_{ij}^\dagger(0)=b_{ji}(0).
\end{equation}
In particular the diagonal components of fermionic zero mode are real and we
will use them later to describe the degeneracy of vacua. Now we concentrate
our attention on non-diagonal zero modes. In the fundamental domain all $z_i$
are different, then one can always make take them to satisfy the inequality
$z_N<z_{N-1}<\dots<z_1$ in this domain. Such condition together with
(\ref{ccr}) leads to interpretation of $a_{ij}(0)$ as creation operator if
$i<j$ and as annihilation operator otherwise. The situation for fermions
is more ambiguous. One can consider $b_{ij}(0)$ as creation operator either
when $i<j$ or when $i>j$, both assumptions are consistent with (\ref{ccr}).
Later we will explore each of these situations.

Let us now discuss the definition of singular operator products
in the current (\ref{current}). We define the current operator
by point splitting:
\begin{equation}
J^+ \equiv \lim_{\epsilon \rightarrow 0}\left( {J^+}_{\phi}(x;
\epsilon)
+{J^+}_{\psi}(x; \epsilon) \right),
\label{current2}
\end{equation}
where the divided pieces are given by
\begin{eqnarray}
&& {J^+}_{\phi}(x; \epsilon)=\frac{1}{i}
\left[{\rm e}^{-i\frac{\pi\epsilon}{2L} M}
\phi(x^- -\epsilon){\rm e}^{i\frac{\pi\epsilon}{2L} M},
D_-\phi(x^-)\right] \label{bcurrent}\\
&& {J^+}_{\psi}(x; \epsilon)=-\frac{1}{\sqrt 2}
\left\{ {\rm e}^{-i\frac{\pi \epsilon}{2L} M}
\psi(x^--\epsilon){\rm e}^{i\frac{\pi\epsilon}{2L} M},
\psi(x^-)\right\}.
\label{fcurrent}
\end{eqnarray}
Here $M$ is diagonal matrix: $M={\mbox diag}(z_1,\dots,z_N)$.
An advantage of this regularization is that
the current transforms covariantly under the
large gauge transformation.

To evaluate (\ref{bcurrent}) and (\ref{fcurrent}) we will generalize
the approach used in \cite{pk96,mrp97} to the SU(N) case. First let
us calculate the vacuum average of bosonic current. Taking into account the
interpretation of zero modes as creation--annihilation operators we obtain:
\begin{eqnarray}
\langle 0|{J^+_{ij}}_{\phi}(x;\epsilon)|0\rangle&=&
\frac{1}{i}\langle 0|{\rm e}^{-i\frac{\pi\epsilon}{2L} (z_i-z_k)}
\phi_{ik}(x^- -\epsilon)D_-\phi(x^-)_{kj}-\nonumber\\
&-&{\rm e}^{-i\frac{\pi\epsilon}{2L} (z_k-z_j)}
\phi_{kj}(x^- -\epsilon)D_-\phi(x^-)_{ik}|0\rangle=\nonumber\\
&=&\frac{1}{4L}\sum_k\sum_{m>0}\left(
{\rm e}^{-i\frac{\pi\epsilon}{2L} (z_i-z_k)}-
{\rm e}^{-i\frac{\pi\epsilon}{2L} (z_k-z_j)}\right){\rm e}^{-ik_m\epsilon}
(\delta_{ij}-\frac{1}{N}\delta_{ik}\delta_{jk})+\nonumber\\
&+&\frac{1}{4L}\sum_{k<j}{\rm e}^{-i\frac{\pi\epsilon}{2L} (z_i-z_k)}
\delta_{ij}-
\frac{1}{4L}\sum_{k>i}{\rm e}^{-i\frac{\pi\epsilon}{2L} (z_k-z_j)}
\delta_{ij}.
\end{eqnarray}
Evaluating the sum and taking the limit one finds:
\begin{equation}
\lim_{\epsilon \rightarrow 0} {J^+_{ij}}_{\phi}(x;\epsilon)
=:{J^+_{ij}}_{\phi}(x): +\frac{1}{4L}\left(z_i-(N+1-2i)\right)\delta_{ij},
\label{bcurrent2}
\end{equation}
where $:{J^+}_{\phi}:$ is the naive normal ordered currents. To be more
precise, we have omitted the zero modes of the diagonal color sectors in
which the notorious constrained zero mode \cite{my76} appears.

The result for fermionic current depends on our interpretation of zero modes
as creation--annihilation operators and it is given by
\begin{equation}
\lim_{\epsilon \rightarrow 0} {J^+_{ij}}_{\psi}(x;\epsilon)
=:{J^+_{ij}}_{\psi}(x): -\frac{1}{4L}\left(z_i\mp(N+1-2i)\right)\delta_{ij}.
\label{fcurrent2}
\end{equation}
The minus sign here corresponds to the case where $b_{ij}(0)$ is a creation
operator if $i<j$ (i.e. the convention is the same as for the bosons) and
plus corresponds to the opposite situation.
As can be seen,
$ {J^+}_{\phi}$ and $ {J^+}_{\psi}$
acquire extra $z$ dependent terms, so called gauge corrections.
Integrating these charges over $x^-$, one finds that the charges are time
dependent. Of course this is an unacceptable situation, and
implies the need to impose
special conditions to single out `physical states' to form
a sensible theory. The
important simplification of the  supersymmetric model is that these time
dependent terms cancel, and the full current  (\ref{current2})
becomes
\begin{equation}
J^+_{ij}(x)=:{J^+_{ij}}_{\phi}:+ :{J^+_{ij}}_{\psi}:+C_i\delta_{ij}.
\label{nocurrent}
\end{equation}
Depending on the convention for fermionic zero modes the {\em z
independent} constants $C_i$ either vanish or they are given by
\begin{equation}
C_i=-\frac{1}{2L}(N+1-2i).
\end{equation}
The regularized current is thus equivalent to the naive
normal ordered current up to an irrelevant constant.
Similarly, one can show that $P^+$ picks up gauge correction when the
adjoint scalar or adjoint fermion are considered separately but in the
supersymmetric theory it is nothing more than  the expected normal ordered
contribution of the matter fields.

In one sense these results are a consequence of the well known fact that the
normal ordering constants in a supersymmetric theory cancel between
fermion and boson contributions. The important point here is that these normal
ordered constants are not actually constants, but rather quantum mechanical
degrees of freedom. It is therefore not obvious that they should
cancel. Of course, this property profoundly effects the
dynamics of the theory.

\section{Vacuum Energy}
The wave function of the vacuum state for the
supersymmetric Yang-Mills theory in 1+1 dimensions
has already been discussed in
the
equal-time formulation \cite{oda95}.
An effective potential is computed in a weak
coupling region as a function of the gauge zero mode
by using the adiabatic  approximation.
Here we analyze the vacuum structure of the same theory
in the context of the DLCQ formulation.

The presence of zero modes renders the light-cone
vacuum quite nontrivial, but the advantage of the light-cone quantization
becomes  evident: the ground state is the Fock vacuum for a fixed
gauge zero mode and therefore our ground state may be
written in the tensor product form
\begin{equation}
\vert \Omega \rangle \equiv \Phi[z]\otimes
\vert 0 \rangle,
\label{vacuum}
\end{equation}
where we have taken the  Schr\"{o}dinger representation
for the quantum mechanical degree of freedom $z$
which is defined in the fundamental domain.
In contrast, to find the ground state of the fermion
and boson for a fixed value of the gauge zero mode
turns out to be a highly nontrivial task in the equal-time
formulation \cite{oda95}.

Our next task is to derive an effective Hamiltonian
acting on $\Phi[z]$. The light-cone Hamiltonian
$H \equiv P^-$ is obtained from energy momentum tensors,
or through the canonical procedure:
\begin{eqnarray}\label{hamil}
H&=&-\frac{g^2L}{4\pi^2}\frac{1}{K(z)}\sum_i
\frac{\partial}{\partial z_i}
K(z)\frac{\partial}{\partial z_i}+ \nonumber\\
&+& \int_{-L}^L dx^-{\rm tr} \left(
-\frac{g^2}{2} J^+\frac{1}{D_-^2} J^+
+\frac{ig^2}{2\sqrt 2}[\phi, \psi]
\frac{1}{D_-}[\phi, \psi]\right),\\
K(z)&=&\prod_{i>j}{\rm sin}^2(\frac{\pi (z_i-z_j)}{2}),
\end{eqnarray}
where the first term is the kinetic energy of the
gauge zero mode,
and in the second term the zero modes of $D_-$ are understood
to be removed. Note that the kinetic term of the gauge
zero mode is not the standard form $-d^2/dz^2$ but acquires a nontrivial
Jacobian $K$ which is nothing but the Haar measure of SU(N).
The Jacobian originates from the unitary transformation of the variable from
$A^+$ to $v$, and can be derived  by explicit evaluation of a functional
determinant \cite{lnt94, lst95}. In the present context
it is found in \cite{kall}. Also we mention that Hamiltonian (\ref{hamil})
seems to contain terms quadratic in diagonal zero modes
$\stackrel{0}{\psi}_{ii}$. However using constraint equations one can show
that the total contribution of all such term vanishes. This also can be
seen by using the fact that Hamiltonian is proportional to the square
of supercharge (\ref{Qminus}).

Projecting the light-cone
Hamiltonian onto the Fock vacuum sector we obtain
the quantum mechanical Hamiltonian
\begin{equation}
H_0=-\frac{g^2L}{4\pi^2}\frac{1}{K(z)}\sum_i
\frac{\partial}{\partial z_i}
K(z)\frac{\partial}{\partial z_i}+V_{JJ}+V_{\phi\psi},
\end{equation}
where the reduced potentials are defined by
\begin{eqnarray}
&& V_{JJ}\equiv -\frac{g^2}{2}\int_{-L}^L dx^-
\langle {\rm tr} J^+\frac{1}{D_-^2} J^+  \rangle, \\
&&V_{\phi\psi}\equiv\frac{ig^2}{2\sqrt 2} \int_{-L}^L dx^-
\langle {\rm tr}
[\phi, \psi]
\frac{1}{D_-}[\phi, \psi]\rangle,
\end{eqnarray}
respectively. As stated in the previous section,
the gauge invariantly regularized current
turns out to be precisely the normal ordered current
in the absence of the zero modes.
It is now straightforward to evaluate $V_{JJ}$ and $V_{\phi\psi}$
in terms of modes. One finds that they cancel among themselves
as expected from the supersymmetry:
\begin{eqnarray}
V_{JJ}&=&-V_{\phi\psi}=\frac{g^2L}{16\pi^2}\left[\sum_{n, m=1}^\infty
\sum_{ijk}\frac{1}{(n-z_i+z_k)(m+z_j-z_k)}-\sum_{n, m=1}^\infty\frac{N}{mn}+
\right.\nonumber\\
&+&\sum_{n=1}^\infty\sum_{ij}\left(\sum_{k>j}\frac{1}{(n-z_i+z_k)(z_j-z_k)}+
\sum_{k<i}\frac{1}{(n+z_j-z_k)(z_k-z_i)}\right)+\nonumber\\
&+&\left.\sum_{ij} \sum_{i>k>j}\frac{1}{(z_k-z_i)(z_j-z_k)}\right].
\label{Vcanc}
\end{eqnarray}
This cancellation was found as the result of formal manipulations with
divergent series like ones in the right hand side of the last formula. Such
transformations are not well defined mathematically and as the result they
may lead to the finite "anomalous" contribution. The famous chiral
anomaly initially was found as the result of careful analysis of
transformations analogous to ones we just performed \cite{adler}. However if
one considers derivatives of $V_{JJ}$ or $V_{\phi\psi}$ with respect to any
$z_i$ then all the sums become convergent, the order of summations becomes
interchangeable and as the result the derivatives of $V_{JJ}+V_{\phi\psi}$
vanish. Thus if there is any anomaly in the expression above it is given by
$z$--independent constant. Such constant in the Hamiltonian would correspond
to the shift of energy levels and usually it is ignored. However in
supersymmetric case there is a natural choice for such constant: in order
for vacuum to be supersymmetric it should be zero. Below we assume that SUSY
is not broken, then we expect that (\ref{Vcanc}) is true.

Thus we arrive at
\begin{equation}
H_0=-\frac{g^2L}{4\pi^2}\frac{1}{K(z)}\sum_i\frac{\partial}{\partial z_i}
K(z)\frac{\partial}{\partial z_i}.
\label{h0}
\end{equation}
The relevant solutions of this equation should be finite in the fundamental
domain, this requirement leads to discrete spectrum due to the fact that
Jacobian vanishes on the boundary of this domain. However the operator $H_0$
is elliptic, and therefore it can't have negative eigenvalues.
If the eigenvalue problem
\begin{equation}
H_0\Phi(z)=E\Phi(z)
\end{equation}
has a solution for $E=0$, this solution corresponds to the ground state
of the theory. It is easy to see that such solution exists and it is given
by $\Phi(z)=const$ \footnote{some authors prefer to rewrite this to
include the measure in the definition of the wave function and then in SU(2)
for example the ground state wave function is a sin}.  We have thus found
that the ground state has a vanishing vacuum energy, suggesting that the
supersymmetry is not broken spontaneously.

\section{Supersymmetry and Degenerate Vacua.}

As we saw in the previous section supersymmetry leads to the cancellation of
the anomaly terms in current operator. However these terms played an
important role in the description of $Z_N$ degeneracy of vacua \cite{pin97a},
so we should find another explanation of this fact here. It appears that
fermionic zero modes give a natural framework for such treatment.

First we will generalize the supersymmetry transformation given in
\cite{sakai95} to the present case, i.e. we include $A^+$ and the zero modes
of fermions. The naive SUSY transformations spoil the
gauge fixing condition, so we combine them with compensating gauge
transformation following \cite{sakai95}. In three dimensional notation
(spinors have two components and indices go from 0 to 2) the result reads:
\begin{eqnarray}\label{SUSYtr}
\delta A_\mu=\frac{i}{2} {\bar\varepsilon}\gamma_\mu \Psi-D_\mu
\frac{i}{2} {\bar\varepsilon}\gamma_-\frac{1}{D_-} \tilde{\Psi},\\
\delta\Psi=\frac{1}{4}F_{\mu\nu}\gamma^{\mu\nu}\varepsilon-\frac{g}{2}[
{\bar\varepsilon}\gamma_-\frac{1}{D_-} \tilde{\Psi},\Psi].\nonumber
\end{eqnarray}
The difference between above expression and those in \cite{sakai95} is that
we include the zero modes. Namely we defined $\Psi$ as the
complete field with all the zero modes included and $\tilde{\Psi}$ as fermion
without diagonal zero modes. The introducing of $\tilde{\Psi}$ is
necessary, because diagonal zero modes form the kernel of operator $D_-$, so
$\frac{1}{D_-}$ is not defined on this subspace.

In particular we are
interested in supersymmetry transformations for $A^+$ and fermionic zero modes.
Performing a mode expansion one can check that diagonal elements of matrix
$[\frac{1}{D_-}\psi,\psi]^0$ vanish, then from (\ref{SUSYtr}) we get:
\begin{eqnarray}\label{zmSUSY}
\delta A^+_{ii}=\frac{i}{\sqrt{2}} {\varepsilon_+^T}\stackrel{0}{\psi}_{ii},
\nonumber\\
\delta \stackrel{0}{\psi}_{ii}=-2\partial_+ A^+_{ii} \varepsilon_+.
\end{eqnarray}
This expression is written in two component notation and the decomposition of
spinor $\varepsilon$: $\varepsilon=(\varepsilon_+,\varepsilon_-)^T$ is used.
Note that since
$\bar{\varepsilon}Q=\sqrt{2}(\varepsilon_+Q^-+\varepsilon_-Q^+)$ the fields
involved in transformations (\ref{zmSUSY}) don't contribute to $Q^+$, this
is consistent to the fact that being $x^-$ independent they don't contribute
to $P^+$. The equations (\ref{zmSUSY}) look like supersymmetry transformation
for the quantum mechanical system built from free bosons and free fermions. In
fact as one can see the supercharge $Q^-$ is the sum of supercharge for the
quantum mechanical system and from the QFT without diagonal zero modes:
\begin{equation}\label{Qminus}
Q^-=-2g\int dx^- \mbox{tr}(J^+\frac{1}{D_-}\psi)+4L\mbox{tr}
    (\partial_+ A^+\stackrel{0}{\psi}).
\end{equation}
Calculating $(Q^-)^2$ and writing the momentum conjugate to $A^+$ as
differential operator \footnote{using Schroedinger coordinate
representation for quantum mechanical degree of freedom - note that the QFT
term has non-trivial dependence on the quantum mechanical coordinate.} we
reproduce Hamiltonian (\ref{hamil}). Note that
$\psi$ there has all the zero modes in it. The square of another supercharge
\begin{equation}
Q^+=2\int dx^-\mbox{tr}(\psi D_-\phi)
\end{equation}
gives $P^+$ while the anti-commutator of $Q^-$ with $Q^+$ is proportional to
the constraint (\ref{fermConstr}) and thus vanishes.

One can check that although $[\stackrel{0}{\psi}_{ii},H]$ does not vanish,
this commutator annihilates Fock vacuum $|0\rangle$, then it also
annihilates $|\Omega\rangle$. In section 2 we mentioned that
$\stackrel{0}{\chi}_{ii}$ decouples from the theory, and therefore it
commutes with
Hamiltonian. Thus acting on the vacuum state $|\Omega\rangle$ by diagonal
elements of either $\stackrel{0}{\psi}$ or $\stackrel{0}{\chi}$ we get states
annihilated by $P^-$ and $P^+$ (the latter statement is obvious since zero
modes commute with momentum). Not all such states however may be considered
as vacua. Although we fixed the gauge in section 2, the theory still has
residual symmetry
$P$, corresponding to permutations of the color basis. Physical states are
constructed from operator acting on the physical vacuum $|\Omega \rangle $
and both
the operators and the physical vacuum must be invariant under
$P$. Such objects can always be written as combinations of
traces. The candidates for the vacuum state may have any combination of
$\stackrel{0}{\psi}$ and $\stackrel{0}{\chi}$ inside the trace, here and
below we consider only diagonal components of zero modes. Since
$\stackrel{0}{\chi}$ is not dynamical we have the usual c--number relation
\begin{equation}
\{\stackrel{0}{\chi}_{ii},\stackrel{0}{\chi}_{jj}\}=0
\end{equation}
instead of canonical anti-commutator, so
$\stackrel{0}{\chi}\stackrel{0}{\chi}=0$. From the relations
(\ref{ccr}) one finds:
\begin{equation}
\stackrel{0}{\psi}\stackrel{0}{\psi}=\frac{1}{4L\sqrt{2}}(1-\frac{1}{N}),
\end{equation}
also we have $\stackrel{0}{\chi}\stackrel{0}{\psi}=-
\stackrel{0}{\psi}\stackrel{0}{\chi}$. Using all these relations and the
$SU(N)$ conditions $\mbox{tr}(\stackrel{0}{\psi})=0$ and
$\mbox{tr}(\stackrel{0}{\chi})=0$ we find that the only nontrivial trace
involving
only zero modes is
$\mbox{tr}(\stackrel{0}{\psi}
\stackrel{0}{\chi})$. Then the family of vacua is given by:
\begin{equation}
\left(\mbox{tr}(\stackrel{0}{\psi}\stackrel{0}{\chi})\right)^n|\Omega\rangle,
\qquad 0\le n\le N-1.
\end{equation}
The region for $n$ is determined taking into account the fact that
$\stackrel{0}{\chi}$ is anti-commuting field with $N-1$ independent
components. Thus we explained the $Z_N$ degeneracy of vacua first mentioned
in \cite{wit79}.

\section{Discussion}
The theory we consider here is an ${\cal N}=1$ super-Yang-Mills theory with one
adjoint fermion and one adjoint scalar with periodic boundary conditions in
$x^-$. The boundary conditions reduce the gauge group to $SU(N)/Z_N$ and
give rise to
diagonal gauge zero modes. We find that supersymmetry requires diagonal
fermion zero
modes which are the supersymmetric partner of the gauge zero mode. 
These zero modes
behave as
quantum mechanical degrees of freedom. When we include
these zero modes in the supercharge we find that  the super-algebra is
unchanged.

In general, when one
normal orders the operators of the theory one finds contributions that depend
only on this quantum mechanical degree of freedom.  These terms are anomalies
and profoundly effect the structure of the theory. In theories with only
fermions or only bosons, these anomalies yield time
dependent charges and a non-zero vacuum energy. In the supersymmetric theory
presented here, these anomalies are seen to cancel and the operators
are all well behaved. In particular, the charges are time independent and
the ground
state is the same as the ground state for the theory without matter. The
energy of
the ground state is zero leaving the supersymmetry unbroken. We show that the
fermion zero modes can be used to construct an $N$-fold set of degenerate
vacua.  This is expected for a theory with the reduced gauge symmetry
$SU(N)/Z_N$.

It is expected that there will be constrained zero
modes which we do not consider here. 
They are not dynamical degrees of freedom
but can
introduce new interactions\footnote{If enough supersymmetry is present,
one expects certain interactions not to be renormalized. 
This would be tantamount to a cancellation of such constrained 
zero-mode degrees of freedom \cite{bas98}.} which 
could lead to supersymmetry breaking
in the same
way that they are known to spontaneously break 
the $Z_2$ symmetry in the simple
$\phi^4_{1+1}$ theory \cite{bpv94}. 

Finally, we remark that the properties of Matrix String Theory
\cite{dvv97} -- which is defined as 1+1 ${\cal N}=8$ super-Yang-Mills
theory on a circle -- depend crucially on the measurable effects produced
by the space-like compactification. These effects are intimately
tied with the dynamics of  non-perturbative objects in Type IIA string theory
known as D0 branes. It would be interesting to consider the
DLCQ formulation of the same Yang-Mills theory, and to establish --
if possible -- any connection with the Matrix String proposal.
The simplicity of the light-cone Fock vacuum, owing to
special supersymmetry cancellations, might present a tractable approach
to non-perturbative string theory.


\medskip
\noindent
\begin{large}
{\bf Acknowledgments}\\
\end{large}
S.T. wishes to thank S.Tanimura for helpful discussions.
S.T. would like to thank Ohio State University for hospitality,
where this work was begun.

%

\vfil

\end{document}